\documentclass[%
 aip,
 amsmath,amssymb,
 preprint,%
]{revtex4-1}

\usepackage{graphicx}
\usepackage{dcolumn}
\usepackage{bm}

\usepackage[utf8]{inputenc}
\usepackage[T1]{fontenc}
\usepackage{mathptmx}
\usepackage{etoolbox}

\newlength{\figurewidth}
\newlength{\figuredoublewidth}

\newcommand{\SIMES}{\affiliation{Stanford Institute for Materials and Energy
Sciences, SLAC National Accelerator Laboratory, 2575 Sand Hill Road, Menlo Park,
California 94025, USA}}
\newcommand{\SCS}{\affiliation{European XFEL, Holzkoppel 4, 22869 Schenefeld,
Germany}}
\newcommand{\LCLS}{\affiliation{Linac Coherent Light Source, SLAC National
Accelerator Laboratory, 2575 Sand Hill Road, Menlo Park, California 94025, USA}}
\newcommand{\DAP}{\affiliation{Department of Applied Physics, Stanford
University, Stanford, California 94305, USA}}
\newcommand{\DP}{\affiliation{Department of Physics, Stanford University,
Stanford, California 94305, USA}}
\newcommand{\UA}{\affiliation{van der Waals-Zeeman Institute, University of
Amsterdam, 1018XE Amsterdam, The Netherlands}}
\newcommand{\ALS}{\affiliation{Advanced Light Source, Lawrence Berkeley National
Laboratory, Berkeley, California 94720, USA}}
\newcommand{\HGST}{\affiliation{San Jose Research Center, HGST a Western Digital
Company, 3403 Yerba Buena Road, San Jose, California 95135, USA}}
\newcommand{\DPS}{\affiliation{Department of Physics, Stockholm University,
Stockholm, 10691 Sweden}}

\newcommand{\TUD}{\affiliation{Department of Applied Physics, Institute for
Photonic Integration, Eindhoven University of Technology, P.O. Box 513, 5600 MB
Eindhoven, The Netherlands}}
\newcommand{\Chem}{\affiliation{Institute of Physics,
Chemnitz University of Technology, D-09107 Chemnitz, Germany}}
\newcommand{\HZDR}{\affiliation{Institute of Ion Beam Physics and Materials
Research, Helmholtz-Zentrum Dresden-Rossendorf, Bautzner Landstrasse 400,
01328 Dresden, Germany}}
\newcommand{\Upp}{\affiliation{Department of Physics and
Astronomy, Uppsala University, Box 516, 75120 Uppsala, Sweden}}

\newcommand{\tspectra}{0.4~ps}
\newcommand{\EF}{E$_\mathrm{F}$}
\newcommand{\CoPd}{[Co\slash{}Pd]}

\makeatletter
\def\@email#1#2{%
 \endgroup
 \patchcmd{\titleblock@produce}
  {\frontmatter@RRAPformat}
  {\frontmatter@RRAPformat{\produce@RRAP{*#1\href{mailto:#2}{#2}}}\frontmatter@RRAPformat}
  {}{}
}%
\makeatother
\begin{document}

\preprint{AIP/123-QED}

\title[State-resolved dynamics in CoPd]{State-resolved ultrafast charge and spin dynamics in \CoPd{} multilayers}


\author{Lo\"ic Le Guyader}
\email{loic.le.guyader@xfel.eu}
\SIMES{}
\SCS{}

\author{Daniel J. Higley}
\SIMES{}
\DAP{}
\LCLS{}

\author{Matteo Pancaldi}
\DPS{}

\author{Tianmin Liu}
\author{Zhao Chen}
\SIMES{}
\DP{}

\author{Tyler Chase}
\SIMES{}
\DAP{}

\author{Patrick W. Granitzka}
\SIMES{}
\UA{}

\author{Giacomo Coslovich}
\author{Alberto A. Lutman}
\author{Georgi L. Dakovski}
\author{William F. Schlotter}
\LCLS{}

\author{Padraic Shafer}
\author{Elke Arenholz}
\ALS{}

\author{Olav Hellwig}
\HGST{}
\Chem{}
\HZDR{}

\author{Mark L.M. Lalieu}
\author{Bert Koopmans}
\TUD{}


\author{Alexander H. Reid}
\SIMES{}
\LCLS{}

\author{Stefano Bonetti}
\DPS{}

\author{Joachim St\"{o}hr}
\SIMES{}

\author{Hermann A. D\"urr}
\email{hermann.durr@physics.uu.se}
\SIMES{}
\Upp{}

\date{\today}

\begin{abstract}
We use transient absorption spectroscopy with circularly polarized x-rays
to detect laser-excited hole states below the Fermi level and compare
their dynamics with that of unoccupied states above the Fermi level in
ferromagnetic \CoPd{} multilayers. While below the Fermi level an
instantaneous and significantly stronger demagnetization is observed,
above the Fermi level the demagnetization is delayed by $35\pm10$~fs.
This provides a direct visualization of how ultrafast demagnetization
proceeds via initial spin-flip scattering of laser-excited holes to the
subsequent formation of spin waves.
\end{abstract}

\maketitle

%

Femtosecond optical excitation of magnetic materials and heterostructures
leads to strongly non-equilibrium conditions displaying many novel
phenomena that are absent in equilibrium physics (for reviews see
Refs.~\onlinecite{Kirilyuk2010} and~\onlinecite{Hellman2017} and
references therein). Discoveries such as all-optical magnetization
reversal~\cite{Stanciu2007, Ostler2012, Mangin2014, Lambert2014},
superdiffusive spin transport~\cite{Battiato2010, Battiato2012, Pfau2012,
Rudolf2012, Eschenlohr2013, Kampfrath2013, Schellekens2014,
Razdolski2017} and optically induced spin transfer 
effect~\cite{Dewhurst2018, Hofherr2020} not only challenge our
fundamental understanding but also provide future perspectives for
information storage and processing. It is generally accepted that laser
excitation initially leads to a highly non-equilibrium electron system
while conserving the total electron spin polarization.\cite{Rhie2003,
Carpene2008, Toews2015} Subsequently, electronic thermalization and
magnetization dynamics set in that distribute the deposited laser energy
over the system’s electron, spin and orbital degrees of freedom and 
ultimately to the lattice.\cite{Koopmans2005, Koopmans2010, Rhie2003,
Carpene2008, Toews2015, Maldonado2020} While such dynamics can be
described by a phenomenological three temperature model, the underlying
physics at play remain hidden behind ad hoc coupling
constants.\cite{Beaurepaire1996}

\begin{figure}
\includegraphics[width=\figurewidth]{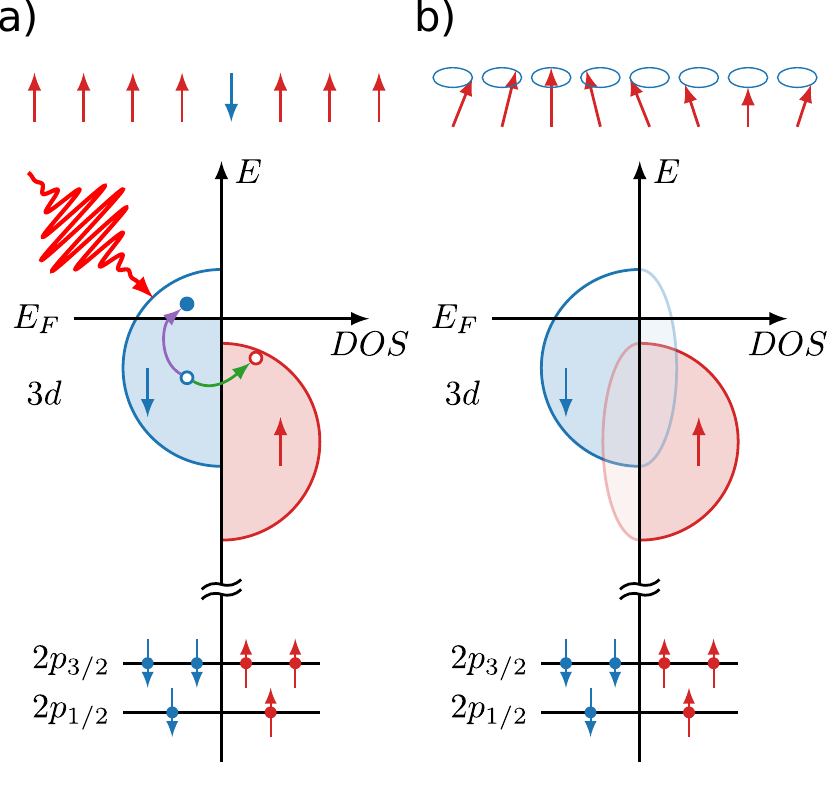}
\caption{\label{fig:Fig1}
(Color online) (a) Schematic of the experiment where the unoccupied $3d$
spin-resolved density of states (DOS) are probed by $2p$ core-level
absorption spectroscopy. Upon excitation by a femtosecond laser pulse,
electrons are promoted from below to above the Fermi level, \EF{}, in a
spin-conserving process (purple arrow). In a strong ferromagnet such as
\CoPd{}, spin relaxation can only occur below \EF{} by a hole spin-flip
(green arrow). (b) After the localized hole spin-flip excitation,
spin-waves are generated and correspondingly the spin-resolved DOS are
partially mirrored.}
\end{figure}

For understanding the angular momentum flow during ultrafast demagnetization it is important to disentangle the influence of magnetic
interactions such as exchange and spin-orbit coupling in the
non-equilibrium dynamics. While the exchange interaction leads to
processes that conserve spin such as angular momentum transfer between
magnetic sub-systems~\cite{Radu2011} and the excitation of spin
waves~\cite{Schaefer2004, Schmidt2010, Goris2011, Turgut2016}, spin-orbit
coupling can cause electronic spin-flips where the corresponding change
in angular momentum is transferred to phonons.\cite{Dornes2019} Such
Elliott--Yafet type spin-flip scattering on phonons has been
investigated and is considered a key ingredient of ultrafast 
demagnetization.\cite{Beaurepaire1996, Koopmans2005, Carva2011, Carva2013} Flipping
electron spins also leads to the excitation of spin
waves~\cite{Vollmer2003} via exchange scattering. However, the direct
observation of spin waves in momentum space and in the time-domain
remains challenging.\cite{Iacocca2019} Several studies reported ultrafast
excitation of spin waves for instance via a delayed onset of
demagnetization.\cite{Mathias2012, Guenther2014, Jana2018} This situation
is visualized in Fig.~\ref{fig:Fig1}(a)
where the initial spin-conserving laser excitation and subsequent
spin-flip processes are depicted. The decay of a flipped electron spin
(top of Fig.~\ref{fig:Fig1}(a)) into spin waves (top of
Fig.~\ref{fig:Fig1}(b)) is  thought to take a characteristic time of
several 10~fs.\cite{Mathias2012, Guenther2014, Jana2018}

Here we use time-resolved x-ray absorption spectroscopy (XAS) and x-ray
magnetic circular dichroism (XMCD) to follow these processes as they
evolve in real time in \CoPd{} multilayers. We show that $2p \to 3d$
core-valence transitions can directly probe the spin-polarization of
laser-induced holes below the Fermi level. These states display an
instantaneous response to fs laser excitation. Surprisingly, $2p \to 3d$
transitions into unoccupied states above the Fermi level show a
demagnetization dynamics that is delayed by $35\pm10$~fs. In addition,
laser-induced hole states below the Fermi level display a much stronger
demagnetization. These observations are consistent with the notion that spin-orbit scattering in
strong ferromagnets is the driving force for ultrafast demagnetization.


Experiments were performed at the SXR instrument of the Linac Coherent
Light Source (LCLS) at the SLAC National Accelerator Laboratory. The
experimental setup is described in detail in
Ref.~\onlinecite{Higley2016}. X-ray absorption spectra were measured
in transmission. The incident X-ray intensity was measured via the x-ray fluorescence from a Si$_3$N$_4$ membrane placed in the beam
before the sample and detected with an microchannel plate (MCP). The
transmitted X-ray intensity behind the sample was recorded by a fast
charge coupled device (CCD) detector. XAS spectra over the L$_3$
absorption edge
corresponding to $2p_{3/2} \to 3d$ transitions were acquired by varying
the x-ray energy via the LCLS electron beam energy. A 250~meV x-ray
bandwidth was selected by the beamline monochromator using a 100
lines per mm grating resulting in an effective resolving power of 3000 at
780~eV\cite{Heimann2011}. Circularly polarized X-ray pulses were produced
using the ``Delta" afterburner undulator\cite{Lutman2016}, enabling the
measurement of XAS and XMCD spectra by alternating the magnetic field
saturating the sample along the beam direction and computing sum and
difference for XAS and XMCD, respectively. Time-resolved XAS and XMCD
data were acquired by scanning the time delay between the 50~fs Full Width at
Half Maximum (FWHM) X-ray probe pulse and the 60~fs FWHM pump laser at a
central wavelength of 798~nm. The data was corrected for timing jitter
between the pump and probe pulses by measuring the arrival time of the
electron pulses via the so-called phase cavity.\cite{Glownia2010} Slower
timing drifts on the few-minutes-scale was corrected using a
cross-correlation-based time delay estimation method as detailed in the
supplementary information. The pump laser was focused on the sample to a
spot size of 190$\times$150~$\mu$m$^2$ FWHM giving a fluence of
$\mathcal{F}$ = 35~mJ/cm$^{2}$. The x-ray spot size was
50$\times$50~$\mu$m$^2$ FWHM and the x-ray fluence below 5 mJ/cm$^{2}$.


A [Co(6\AA)\slash Pd(6\AA)]$_{38}$ multilayer sample capped with a
Pd(20\AA) layer and grown onto a 100~nm Si$_3$N$_4$ membrane with a
Ta(10\AA)\slash Pd(30\AA) buffer layer was used in the measurements
described below. The sample was grown by DC magnetron sputtering with
fabrication details given in the supplementary information. Prior to the
LCLS experiments the sample was characterized at beamline 4.0.2 of the
Advanced Light Source (ALS) using XAS and XMCD measurements, where sum
rules analysis confirmed the magnetic properties of the multilayer with
previously published work as discussed in the supplementary information.


Conceptually, the experiment is depicted in the schematic shown in
Fig.~\ref{fig:Fig1}. In an itinerant strong ferromagnet such as Co in
[Co/Pd] multilayers, the density of states (DOS) can be separated into
completely occupied majority (spin ``up") and partially occupied minority
spin (spin ``down") channels which are shifted in energy by
the exchange splitting. At the L$_3$ absorption edge, valence hole states
are probed via $2p_{3/2} \to 3d$ core-valence transitions. In the ground
state all electronic states up to the Fermi level \EF{} are occupied. As
the pump laser pulse excites the $3d$ electronic system by promoting
electrons from below to above the Fermi level, transient XAS can detect
the additional hole states below \EF{}. XAS transitions into states above
\EF{}, however, are reduced by the laser-excited transient electron
population in these states. Exactly at \EF{} no XAS changes should be
observed. Laser-excited holes below \EF{} are thought to lead to
demagnetization in strong ferromagnets via spin-flip scattering
events,\cite{Carva2011, Carva2013} where an electron from the majority spin fills
the hole in the minority spin, as depicted in Fig.~\ref{fig:Fig1}(a).
This flipped spin could then decay into spin waves as illustrated in
Fig.~\ref{fig:Fig1}(b), which induces a band mirroring in the nearby
atoms where the quantization axis has now changed. By using time-resolved
XAS and XMCD we aim at uncovering the different timescales and energies
of the different processes involved.

\begin{figure}
\includegraphics[width=\figurewidth]{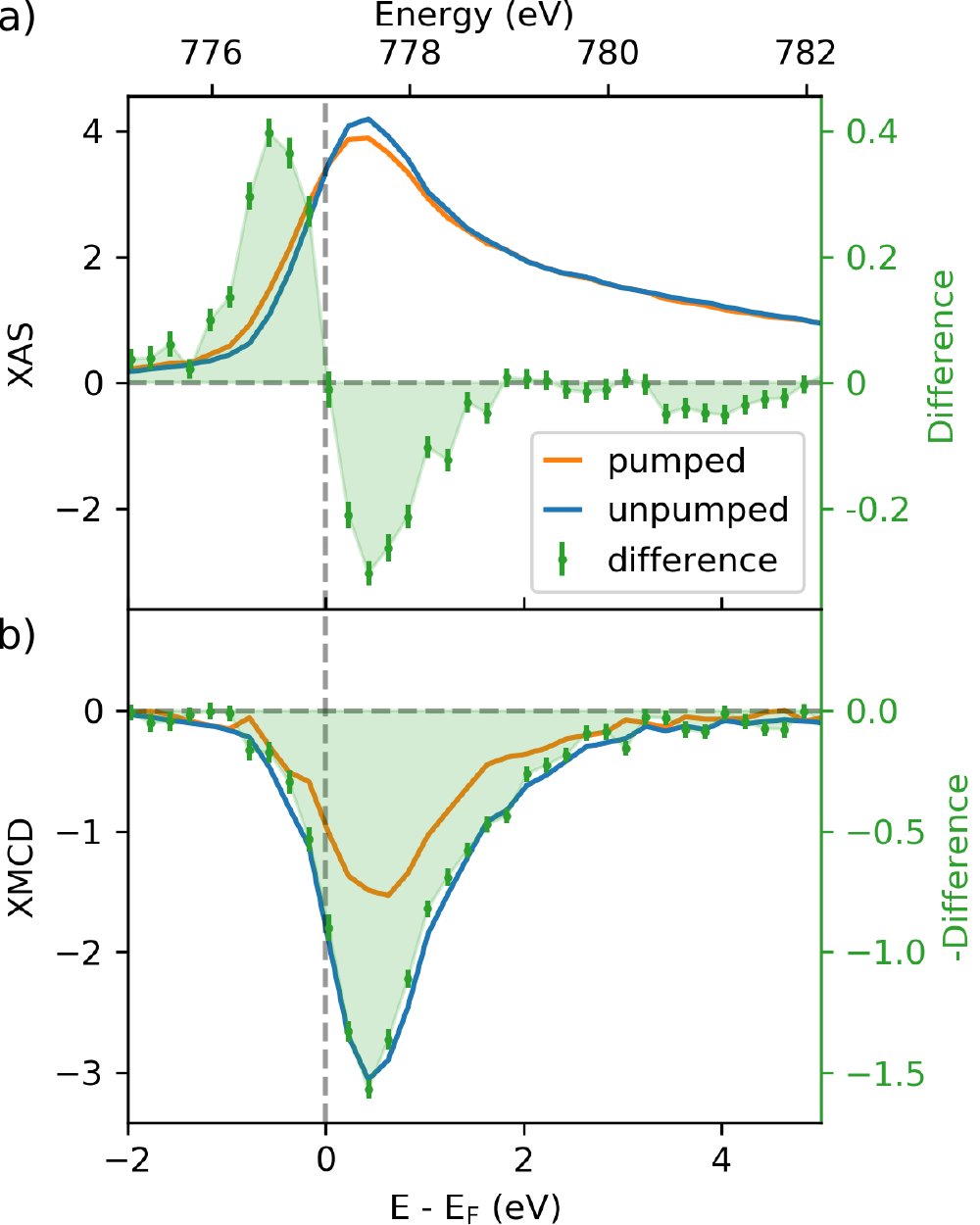}
\caption{\label{fig:dXAS-spectra}
(Color online) Pumped, unpumped and their differences in (a) XAS and (b)
XMCD at the Co L$_3$ edge at a delay of \tspectra{}. The vertical dashed
line indicates the position of the observed zero crossing at the Fermi
level \EF{}. The photon energy is shown on the top axis, while the
energy with respect to the Fermi level is shown on the bottom. The
differences (pumped-unpumped) are shown on a separate vertical axis on
the right. For the XMCD difference the sign was reversed to ease visual
comparison with the unpumped XMCD profile.
}
\end{figure}

Fig.~\ref{fig:dXAS-spectra} shows transient XAS and XMCD of laser-induced holes belwo and above the Fermi level.
XAS (Fig.~\ref{fig:dXAS-spectra}(a)) and XMCD spectra
(Fig.~\ref{fig:dXAS-spectra}(b)) were measured at a fixed time delay of
\tspectra{} at the Co L$_3$ edge. The pump induced changes are shown as
green symbols and shading. While the change in XMCD appears to be mostly
an homogeneous reduction at all photon energies, the change in XAS
clearly displays a derivative-like shape with a zero crossing at an x-ray
energy of 777.2~eV (see top axis of Fig.~\ref{fig:dXAS-spectra}) as 
indicated by the dashed vertical line. At lower x-ray energy the XAS
signal is increased as expected for fs laser-induced hole states. At
higher energy XAS transitions into previously unoccupied states are
blocked by laser-excited electrons leading to the observed intensity
reduction. It is, therefore, possible to identify 777.2~eV as the
position of the Fermi level (see bottom axis of
Fig.~\ref{fig:dXAS-spectra}).\cite{Oppeneer2004}

\begin{figure}
\includegraphics[width=\figurewidth]{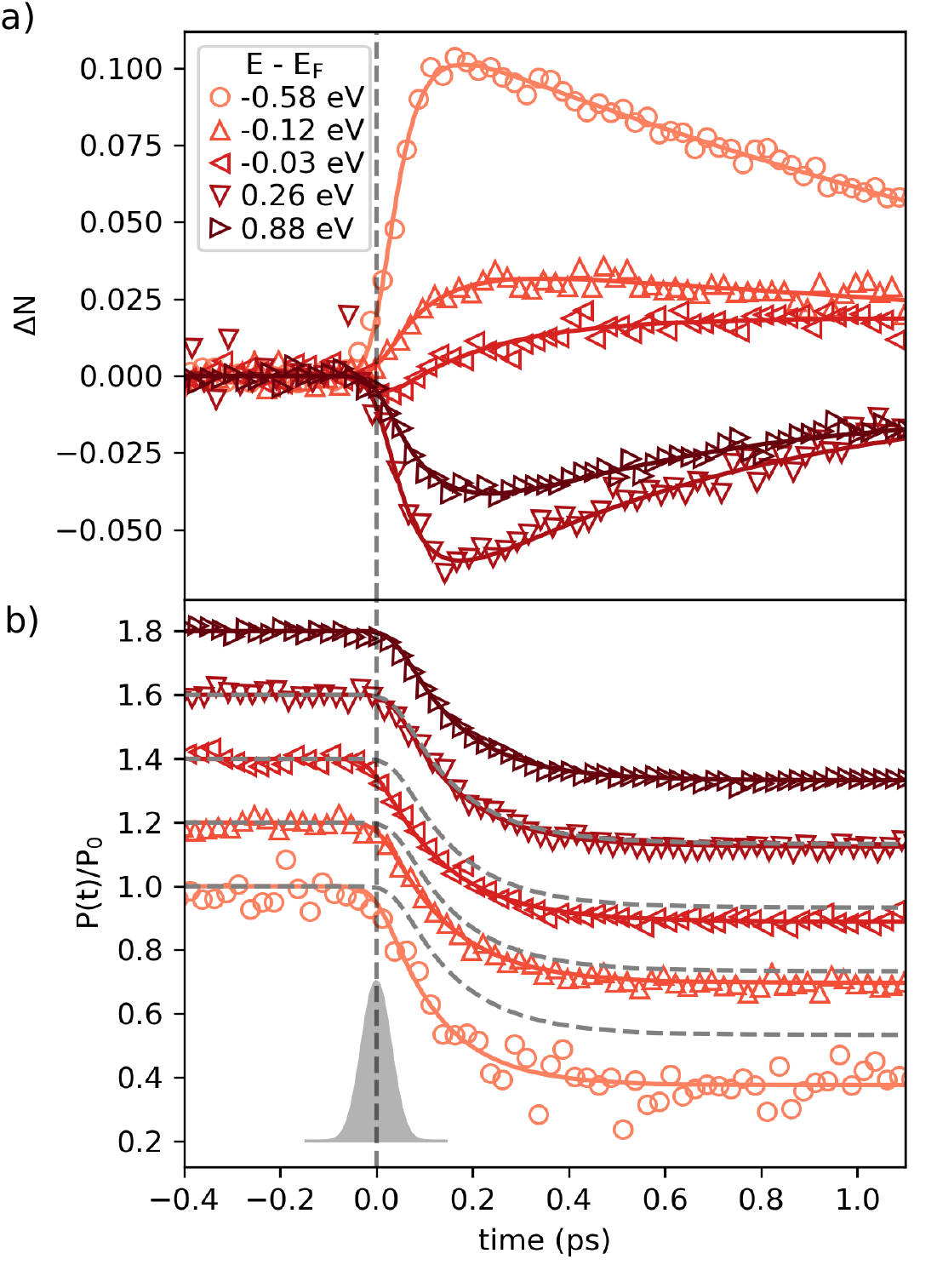}
\caption{\label{fig:L3waterfall}
(Color online) Time-resolved change in state resolved (a) charge
$\Delta$N and (b) relative polarization change P(t)/P$_0$ around the
\EF{} at the L$_3$ edge. In (b), the data are shifted vertically for
clarity and the gray dashed curves are the fit as explained in the text
at E-\EF{} = 0.88~eV.
}
\end{figure}

Fig.~\ref{fig:L3waterfall} displays time-delay traces obtained for
various state energies relative to the Fermi level, E-\EF{}. Below \EF{}
the curves display initial increases in the XAS intensity followed by
subsequent decays on timescales longer than several 100~fs. Above \EF{}
the transient XAS changes are negative while directly at \EF{} a more
complex behavior emerges. In the following we describe these observations
in terms of changes in the hole population, $\Delta$N, at state of energy
E-\EF{}. The small contribution due to spin-orbit coupling of the
state-resolved XAS intensity \cite{Wu1994,Ebert1996} will be neglected
here. It is important to emphasise that $\Delta$N can also include time
dependent changes in the electronic structure.\cite{Stamm2007} It is
apparent in Fig.~\ref{fig:dXAS-spectra}(a) that such electronic structure
changes indeed occur. For instance at E-\EF{} near 4~eV, i.e. much higher
than the pump photon energy, the observed variations of $\Delta$N are
unlikely to be caused by the population dynamics of electrons in these states.

The curves for $\Delta$N in Fig.~\ref{fig:L3waterfall}(a) were fitted
with a double exponential to describe an excitation and a relaxation
process. The fit parameters are summarized in Table II in the
supplementary information. Far above and below the Fermi level, the
initial rise times of $\Delta$N are essentially determined by the length
of the pump pulses. The subsequent decay time scales are shorter further
away from the Fermi level as one would expect from a Fermi liquid
behavior of the electronic system.

The XMCD spectra can be described as being proportional to the product of
state-dependent population, $N$ and a polarization term, $P$. The latter
contains both spin and orbital polarization with the orbital contribution
being significantly smaller than the spin polarization as shown in the
sum rule analysis detailed in the supplementary
information.\cite{Wu1994,Ebert1996} Similar to conventional
sum-rule-analysis of time-resolved XMCD spectra, the magnetic dipole term
can be neglected for our poly-crystalline samples
\cite{Stamm2010,Boeglin2010} and was found to be negligible in similar
samples.\cite{Guo1995} Using the results for $\Delta$N from
Fig.~\ref{fig:L3waterfall}(a) we can separate the state polarization from
state-dependent charge dynamics. The time-resolved polarization dynamics,
normalized to the ground-state polarization of the respective
states, are shown in Fig.~\ref{fig:L3waterfall}(b). The individual
experimental results (symbols) are shifted vertically for clarity and are
compared to exponential polarization decays that include the magnitude of
the decay, $\Delta$P, the decay time constant, $\tau$, and a delayed
demagnetization onset, $\Delta$t, as fit parameters. All parameters are
summarized in Table II in the supplementary information. To highlight
the difference between the curves below and above the Fermi level, the 
same relative polarization dynamics for E-\EF{} = 0.88~eV is shown as a
dashed
grey curve together with each trace. This allows two visual observations.
Firstly, the amount of demagnetization is not the same at each value of
E-\EF{}.  Clearly, the demagnetisation is significantly stronger below
\EF{} than above. Secondly, there is a time delay, $\Delta$t, apparent in
the response, with faster dynamics for states below \EF{}. The complete
fitting model and analysis of the uncertainties on the fitted values is
presented in Table II in the supplementary information. In
Fig.~\ref{fig:t0} the relative change in polarization $\Delta$P/P$_0$ and
the time lag $\Delta$t are shown as function of E-\EF{}. We stress here that this time lag $\Delta$t is determined by the delayed apparent
response of the relative change in polarization $\Delta$P/P$_0$ with respect to the charge dynamics $\Delta$N, for the same given photon energy. It can also be visualized by comparing the relative change in polarization $\Delta$P/P$_0$ at different photon energy but this is not how we extracted it.

\begin{figure}
\includegraphics[width=\figurewidth]{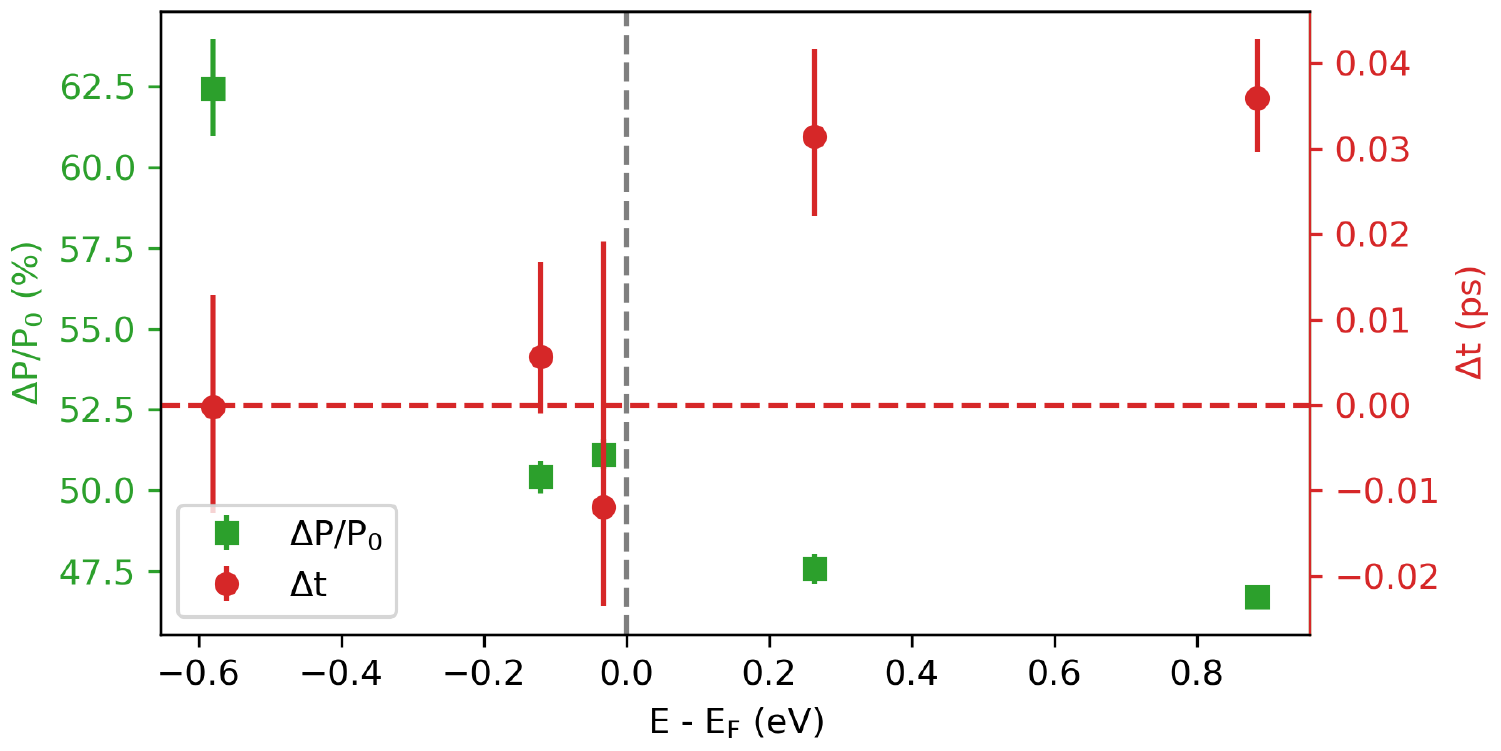}
\caption{\label{fig:t0}
(Color online) Fitted relative change in polarization $\Delta$P/P$^0$ and
time lag $\Delta$t in polarization change response as function of E-\EF{}
at L$_3$ edge.
}
\end{figure}

The data we presented in this letter conclusively demonstrate a vastly
different magnetization dynamics above and below the Fermi level for Co
$3d$ levels in \CoPd{} multilayers. Below \EF{}, the ultrafast drop in
magnetic polarization is up to 32$\%$ larger than above (see
Fig.~\ref{fig:t0}). This is clearly outside any experimental uncertainty
as demonstrated in Fig.~\ref{fig:L3waterfall}(b). Moreover, the onset of the
polarization dynamics occurs simultaneously to the charge dynamics, i.e.
$\Delta$t = $0\pm10$~fs. This is the behavior expected for individual
electrons/holes being scattered between different electronic states as
depicted in Fig.~\ref{fig:Fig1}(a). This also leads to Stoner excitations
where electrons/holes are scattered between the spin up and down
states.\cite{Turgut2016} In strong ferromagnets such as \CoPd{}
multilayers, spin-flip scattering can only occur for states below \EF{}
where spin up and down states are hybridized via spin-orbit
coupling.\cite{Carva2013} The same Elliot-Yaffet-type spin-flip
scattering processes are thought to also transfer spin angular momentum
to the lattice.\cite{Carva2011, Carva2013, Dornes2019}

However, for ultrafast demagnetization to occur, the flipped spins of
individual electrons/holes need to be transferred to the whole electronic
system. This usually takes place via the formation of collective spin
excitations, i.e. spin waves. In the Heisenberg model, spin waves lead to
slight changes of the atomic spin quantization axis and result in a
mixing of spin up and down states as observed in photoemission
spectroscopy.\cite{Eich2017} This situation is depicted in
Fig.~\ref{fig:Fig1}(b). Since the formation of spin waves takes
time~\cite{Mathias2012} we expect a characteristic time delay relative to
the instantaneous demagnetization of individual electrons/holes. We
assign the observed delayed onset of demagnetization above \EF{} of
$\Delta$t = $35\pm10$~fs (see Fig.~\ref{fig:t0}) to this effect. This is
observable above the Fermi level since there the majority of unoccupied
states reflects the atomic magnetic moments.\cite{Carra1993, Thole1992}


In summary, taking advantage of the high FEL brightness and improved
I$_0$ normalization scheme, we were able to show that time-resolved XAS
and XMCD spectroscopy can provide detailed information in the microscopic
mechanism at play during ultrafast laser excitation. In particular we
report on different dynamics of the spin system below and above the Fermi
level. This is manifested by both a 32\% larger change in spin dynamics
below the Fermi level and a $35\pm10$~fs delayed response above it. Both
of these effects suggest a scenario for a strong ferromagnet where
spin-flips occur preferentially below the Fermi level where spin up and
down states are hybridized. Moreover, we also report on initial evidence
that indicates effects beyond a simple electronic redistribution and
demagnetization with changes in XAS observed 4~eV above the Fermi level,
suggestive of band structure dynamics. With ever improved normalization
schemes and higher repetition rate FELs, transient near-edge soft X-ray
spectroscopy promises to be a valuable tool in understanding
out-of-equilibrium phenomena.

\section*{Supplementary Material}

See supplementary material for the sample preparation, sum-rules
analysis, timing drift correction, pump and probe absorption profiles
in the sample and finally the complete fitting model with
uncertainty estimation.

\begin{acknowledgments}
L.L.G. acknowledges the Volkswagen-Stiftung for the financial
support through the Peter-Paul-Ewald Fellowship.
Work at SLAC and the operation of LCLS are supported by the U.S. Department of Energy, Office of Science.
\end{acknowledgments}

\section*{Data Availability Statement}

The data that support the findings of this study are available from the corresponding
author upon reasonable request.

\bibliography{biblio}

\end{document}